\documentclass{arxiv_tlp}
\usepackage{mathptmx}

\usepackage{url}
\usepackage[pdftex]{graphicx,color}
\usepackage{amsmath}
\usepackage{amssymb}
\usepackage[mathscr]{euscript}
\usepackage{stmaryrd}
\usepackage{bussproofs}
\usepackage{pifont}
\usepackage{multirow}
\usepackage{enumitem}
\usepackage{pbox, ragged2e}
\usepackage{fancyvrb}
\usepackage{color}
\usepackage{subfig}
\usepackage{adjustbox}


\DefineVerbatimEnvironment{egcode}{Verbatim}{}

\newtheorem{definition}{Definition}

\DeclareMathAlphabet{\mathcal}{OMS}{cmsy}{m}{n}

\newcommand{\qed}[0]{\square}

\newcommand{\HH}[0]{\mathcal{H}}
\newcommand{\DD}[0]{\mathcal{D}}

\newcommand{\triple}[3]{\{#1\}~#2~\{#3\}}

\newcommand{\false}[0]{\mathit{false}}
\newcommand{\dom}[0]{\mathit{dom}}
\newcommand{\eqdef}[0]{\overset{\mathsf{def}}{=}}
\newcommand{\set}[1]{\mathsf{#1}}

\newcommand{\prop}[0]{\Longrightarrow}

\newcommand{\heap}[0]{{\mathcal{H}}}

\newcommand{\hone}[2]{#1 \mapsto #2}
\newcommand{\hsep}[0]{{\symbol{42}}}
\newcommand{\heq}[0]{\bumpeq}
\newcommand{\hin}[3]{(#2, #3)\in#1}

\newcommand{\foot}[0]{\mathit{Fp}}
\newcommand{\fram}[0]{\mathit{Cxt}}

\newcommand{\ignore}[1]{}

\newcommand{\cmark}[0]{\ding{51}}
\newcommand{\xmark}[0]{\ding{55}}

\newcommand{\RN}[1]{\uppercase\expandafter{\romannumeral #1\relax}}

\newcommand{\authnote}[2]{{\bf \textcolor{blue}{#1}: \em \textcolor{red}{#2}}}
\renewcommand{\authnote}[2]{}

\newcommand{\mysubsubsection}[1]{\vspace{0.25em}\noindent\textbf{#1.}}

\begin{document}

\hyphenation{either}
\providecommand\AMSLaTeX{AMS\,\LaTeX}
\newcommand\eg{\emph{e.g.}\ }
\newcommand\etc{\emph{etc.}}
\newcommand\bcmdtab{\noindent\bgroup\tabcolsep=0pt%
  \begin{tabular}{@{}p{10pc}@{}p{20pc}@{}}}
\newcommand\ecmdtab{\end{tabular}\egroup}
\newcommand\rch[1]{$\longrightarrow\rlap{$#1$}$\hspace{1em}}
\newcommand\lra{\ensuremath{\quad\longrightarrow\quad}}

\title[Shape Neutral Data-structure Analysis]
    {Shape Neutral Analysis of Graph-based Data-structures}
\author[Duck, Jaffar, and Yap]
       {GREGORY J. DUCK and JOXAN JAFFAR and ROLAND H. C. YAP\\
        Department of Computer Science, National University of Singapore\\
        \email{\{gregory,joxan,ryap\}@comp.nus.edu.sg}}

\maketitle

\begin{abstract}
Malformed data-structures can lead to runtime errors such as
arbitrary memory access or corruption.
Despite this, reasoning over data-structure properties for low-level
heap manipulating programs remains challenging.
In this paper we present a constraint-based program analysis that checks
data-structure integrity, w.r.t. given target data-structure properties,
as the heap is manipulated by the program.
Our approach is to automatically generate a solver for 
properties using the type definitions from the target program.
The generated solver is implemented using a
Constraint Handling Rules (CHR) extension of built-in heap, integer and
equality solvers.
A key property of our program analysis is that the target data-structure
properties are \emph{shape neutral}, i.e., the analysis does not
check for properties relating to a given data-structure
graph \emph{shape}, such as doubly-linked-lists versus trees.
Nevertheless, the analysis can detect errors in a wide range of
data-structure manipulating programs, including those that use
lists, trees, DAGs, graphs, etc.
We present an implementation that uses
the Satisfiability Modulo Constraint Handling Rules (SMCHR) system.
Experimental results show that our approach works well for real-world
C programs.
\end{abstract}

\begin{keywords}
Constraint Handling Rules,
Satisfiability Modulo Constraint Handling Rules,
Satisfiability Modulo Theories,
Program Analysis,
Data-structures,
Memory Errors
\end{keywords}

\section{Introduction}\label{sec:intro}

Low-level languages such as 
C and C++ are notorious for (subtle) bugs due to direct pointer
manipulation.
Program analysis may detect bugs,
however, automating such analysis for data-structure manipulating programs
in low-level languages is a challenging problem.
Much of the existing work on data-structure
analysis~\cite{berdine07shape,berdine05smallfoot,berdine11slayer,dudka11predator} focuses on
(or depends on)
\emph{shape} properties, i.e., is the data-structure a tree or linked-lists,
etc.?
However, this complicates automated analysis:
\begin{enumerate}
    \item Shape information is usually \underline{\emph{implicit}}.
    \item Common data-structure shapes have
        \underline{\emph{inductive}} (a.k.a. \emph{recursive}) definitions.
    \item Data-structure integrity co-depends on
        \underline{\emph{memory safety}}.
\end{enumerate}
For example, consider the following generic C \verb+struct+ declaration:
\begin{egcode}
        struct node { node *next1; node *next2; ...};
\end{egcode}
Automated shape discovery on type declarations alone is not feasible:
this node could be for a tree, DAG, graph, doubly-linked-list, etc.
Even if shape information were available (or assumed), the next
problem is that data-structure shapes have
\emph{in\-duct\-ive} (a.k.a. \emph{re\-curs\-ive})
definitions, further complicating automated reasoning.
For example, a list can be recursively defined as
follows (in Separation Logic~\cite{reynolds02separation}):
$\mathsf{list}(l) \eqdef
        \exists t: l = 0 \lor l \mapsto t ~\hsep~ \mathsf{list}(t)$.
Finally, data-structure reasoning for low-level programs
is (co)dependent on \emph{memory safety}, e.g., an object bounds
error may clobber memory that invalidates the data-structure invariant.
Conversely data-structure invariant violations may give rise to memory errors.

In this paper, we present a \emph{shape neutral} data-structure
analysis that aims to avoid 
the complications listed above.
Instead of
data-structure shape definitions, we analyze the
program against a set of more general data-structure properties
that hold for any canonical graph-based data-structures in 
standard idiomatic C.
The properties include:
\begin{enumerate}
\item \label{case:111}
    nodes are contiguous regions of memory.
\item nodes are the correct size.
\item \label{case:222} nodes form a closed directed graph (no dangling links).
\item \label{case:333} nodes do not overlap with each other.
\end{enumerate}
Such general properties eliminate the need to infer (or assume) data-structure
shapes, and can be derived solely from the types declared by the program.

Our automated shape neutral data-structure
analysis for C programs is based on: (1) \emph{symbolic execution} to
generate path constraints for all possible paths through the program,
and (2) a specialized \emph{constraint solver} for shape neutral data-structure
properties.
For the latter, we first formalize the properties
we aim to enforce, then use our formalization to derive a solver that
can be implemented using \emph{Constraint Handling Rules}
(CHR)~\cite{fruhwirth98theory,fruhwirth09chr}.
Since the generated \emph{verification conditions} (VCs) typically
have a rich structure (quantifiers, conjunction, disjunction and negation),
we implement the solver using the \emph{Satisfiability Modulo Constraint
Handling Rules} (SMCHR) system~\cite{duck12smchr,duck13smchr}.
We demonstrate that our implementation is able to verify many data-structure
manipulating
C programs, including data-structures for lists, trees, DAGS, graphs, etc.
We also compare against related tools for finding memory errors
related to violations of the data-structure integrity constraints.

In summary, the main contributions of this paper are:
\begin{enumerate}
\item \underline{\emph{Data-structure Integrity Constraints}}:
    We propose and formalize a set of shape neutral
    data-structure integrity constraints
    based on the properties informally described above.
    The integrity constraints cover standard idiomatic C
    graph-based data-structures.

\item \underline{\emph{Constraint Handling Rules Implementation of Shape
        Neutral Data-structure Properties}}: \\
    We also present a constraint solver for the integrity constraints
    that can be implemented using a Constraint Handling Rules (CHR) extension
    of built-in heap, integer and equality solvers.
    The CHR solver is automatically generated from the target program's
    data-structure type declarations.
    Goals generated by program analysis can then be solved using the
    Satisfiability Modulo Constraint Handling Rules (SMCHR) system.
    SMCHR is suitable as
    it can efficiently handle goals
    with a complex Boolean structure, including negation.
    Furthermore, SMCHR allows different types of solvers (integer, heap,
    data-structure) to be seamlessly integrated.
    
\item \underline{\emph{Evaluation}}:
    Finally we present an experimental evaluation of our overall approach.
    We show that the proposed method is effective on ``real world''
    data-structure manipulating code such as that from the \texttt{GNU GLib}
    library.
    We also compare our approach against several existing
    state-of-the-art memory analysis/safety
    tools.
    We demonstrate that our tool
    can detect memory errors missed by other systems---especially
    regarding more complex data-structures
    involving multiple node types and sharing.
\end{enumerate}

%%%%%%%%%%%%%%%%%%%%%%%%%%%%%%%%%%%%%%%%%%%%%%%%%%%%%%%%%%%%%%%%%%%%%%%%%%%%%%
\section{Preliminaries}\label{sec:prelim}

\begin{figure}[t,fragile]
\subfloat[Summary of the symbolic execution rules.\label{fig:rules}]{
\small
\begin{minipage}{0.5\textwidth}
\begin{align*}
\mathit{sp}((s_1; s_2), \phi) ~\eqdef~ & 
    \mathit{sp}(s_2, \mathit{sp}(s_1, \phi)) \\
\mathit{sp}(x := E, \phi) ~\eqdef~ & 
    x = E[x'/x] \land \phi[x'/x] \\
\mathit{sp}(x := *p, \phi) ~\eqdef~ &
    \mathsf{access}(\heap, p, x) \land \phi[x'/x] \\ 
\mathit{sp}(*p := E, \phi) ~\eqdef~ &
    \mathsf{assign}(H', p, E, \heap) \land \phi[H'/\heap] \\
\mathit{sp}(p := \mathsf{malloc}(n), \phi) ~\eqdef~ &
    \mathsf{alloc}(H', p, n, \heap) \land \phi[H'/\heap, p'/p] \\
\mathit{sp}(\mathsf{abort}(), \phi) ~\eqdef~ & \false \\
\end{align*}
\end{minipage}
}
~~~~
\vline
\subfloat[Program with loops.\label{fig:loops}]{
\begin{minipage}{0.3\textwidth}
\begin{align*}
\begin{array}{l}
(0)\texttt{ S1;}\\
(1)\texttt{ while (b1) \{} \\
~~~~(2)\texttt{ S2;} \\
~~~~(3)\texttt{ while (b2)} \\
~~~~ ~~~~(4)\texttt{ S3;} \\
~~~~(5)\texttt{ S4; \}} \\
(6)~\texttt{ S5;}~(7) \\
\\
\\
\\
\end{array}
\end{align*}
\vspace{0.15em}
\end{minipage}
}
\caption{Strongest post-condition semantics and an example program.}
\end{figure}

Our analysis is based on defining 
a \emph{data structure integrity constraint} (DSIC).
The DSIC formula is derived from a schema which, when given type declarations,
is instantiated into a first-order formula ${\cal D}$.
Essentially, ${\cal D}$ states that data structures must
have valid nodes, valid pointers, and nodes do not intersect.
The formal presentation of ${\cal D}$ is given in Section
\ref{sec:closed}.

The framework we use to analyze a program in pursuit of our
integrity constraint ${\cal D}$ is a classic one:
\emph{Verification Condition Generation} (VCG) via \emph{symbolic execution},
a method originating from Floyd (see e.g. \cite{matthews06lpar}
for a succinct introduction).  The overall algorithm is summarized as follows:
\begin{itemize}
\item[-] The program, interpreted as a graph, is annotated with ${\cal D}$ at certain program points
corresponding to a set of \emph{cut-points} in the control flow graph.
\item[-] VCG is performed as follows. Suppose ${\cal D}$
holds when control reaches some cut-point $p$, then let $q$ be the next subsequent cut-point encountered
during program execution.
We then show that $q$ also satisfies ${\cal D}$.
This is repeated for all cut-points.
\end{itemize}

\noindent
This reduces
shape neutral data-structure analysis
(abbr. to $\DD$-analysis) into proving that \emph{Hoare triples}
of the form $\triple{\DD}{C}{\DD}$ are valid, where $C$ is some
code fragment (the analysis target)
and $\DD$ is the desired DSIC, defined later.
Intuitively, a Hoare triple $\triple{A}{C}{B}$ states that if $A$ holds before
execution of $C$, then $B$ must hold after.
Thus, $\triple{\DD}{C}{\DD}$ is stating that the DSIC
$\DD$ is preserved by $C$.
For branch-free $C$, our underlying methodology is
\emph{symbolic execution} as defined
by the \emph{strongest post condition} (SPC)
\emph{predicate transformer semantics} shown in Figure~\ref{fig:rules}.
All dashed variables (e.g. $x'$) are implicitly existentially quantified,
and the notation $\phi[x'/x]$ represents formula $\phi$ with variable
$x'$ substituted for variable $x$.
We assume the standard definitions for sequences $(s_1; s_2)$,
assignment $x := E$ and $\mathsf{abort}()$.
Heap operations use special heap constraints defined below.
A triple $\triple{\DD}{C}{\DD}$ is established by symbolically executing $\DD$
through $C$ using the rules from Figure~\ref{fig:rules}.
This process generates a \emph{path constraint} $P$.
The triple holds iff \emph{Verification Condition} (VC) $(P \models \DD)$
is proven \emph{valid}
with the help of a suitable constraint solver.

Cut-points are chosen to break loops into straight line program
fragments amenable to symbolic execution.
For example, consider the program with
nested loops shown in Figure~\ref{fig:loops}.
Also consider the formula $\DD$ which we use to annotate points $(0)$, $(1)$,
$(3)$ and $(7)$ (i.e., the chosen ``cut-points'').
The Hoare triples of interest are therefore:
\begin{align*}
\begin{array}{c}
\triple{\DD}{\mathtt{S1}}{\DD} ~~~~~~
\triple{\DD \wedge \mathtt{b1}}{\mathtt{S2}}{\DD} ~~~~~~
\triple{\DD \wedge \mathtt{b2}}{\mathtt{S3}}{\DD} \\
\triple{\DD \wedge \neg \mathtt{b2}}{\mathtt{S4}}{\DD} ~~~~~~
\triple{\DD \wedge \neg \mathtt{b1}}{\mathtt{S5}}{\DD}
\end{array}
\end{align*}

\noindent
It is important to note that we are not requiring integrity at \emph{every}
program point,
rather only at the cut-points, which is a heuristic that works reasonably well
in practice (see Section~\ref{sec:experiments}).

\mysubsubsection{Heap Operations}
To handle heap operations, we extend the $\HH$-constraint language
from~\cite{duck13heaps}.
We assume, as given, a set of
$\set{Values}$ (typically $\set{Values} \eqdef \mathbb{Z}$)
and define the set of $\set{Heaps}$ to be all
\emph{finite partial map}s between values, i.e.,
$\set{Heaps} \eqdef (\set{Values} \rightharpoonup_{\text{fin}} \set{Values})$.
Let $\dom(H)$ be the \emph{domain} of the heap $H$.
We abuse notation and treat heaps $H$ as sets of (pointer,value)
pairs $\{(p, H(p))~|~p \in \dom(H)\}$.
Conversely, a set of pairs $S$ is a \emph{heap} iff for all $p, v, w$
we have that 
$(p, v), (p, w) \in S \rightarrow v = w$.
A \emph{heap partitioning constraint} is a formula of the form
$H \heq H_1 \hsep H_2$, where $H, H_1, H_2$ are heap variables.
Informally, the constraint $H \heq H_1 \hsep H_2$ states that
heap $H$ can be partitioned into two disjoint (separate) sub-heaps
$H_1$ and $H_2$.
The set-equivalent definition is as follows:
$H = H_1 \cup H_2 \land \dom(H_1) \cap \dom(H_2) = \emptyset$.

We use the symbolic execution rules for heap operations
from~\cite{duck13heaps} summarized in
Figure~\ref{fig:rules}.
By convention, the state of the program heap is represented by a
\emph{distinguished heap variable} $\heap$ (of type $\set{Heaps}$).
Each heap operation modifies $\heap$ according to some
heap constraint $\mathsf{access}$, $\mathsf{assign}$, and
$\mathsf{alloc}$ defined as follows:
\begin{align*}
 &   \mathsf{access}(\heap, p, v) \eqdef \hin{\heap}{p}{v} \\
 &   \mathsf{assign}(H, p, v, \heap) \eqdef 
    \exists w: 
        \hin{H}{p}{w} \land
        \heap = (H - \{(p, w)\}) \cup \{(p, v)\} \\
 &   \mathsf{alloc}(H, p, 1, \heap) \eqdef 
        \exists w: \heap = H \cup \{(p, w)\} \land p \not\in \dom(H)
\end{align*}
We can extend the definition for arbitrary-sized $\mathsf{alloc}$ in
the obvious way.
Note that our definitions implicitly assume that accessing
\emph{unmapped memory} (i.e. any $p \not\in \dom(\heap)$)  behaves the same
way as $\mathsf{abort}()$ (see Figure~\ref{fig:rules}).

%%%%%%%%%%%%%%%%%%%%%%%%%%%%%%%%%%%%%%%%%%%%%%%
\section{Data-Structure Analysis}\label{sec:closed}

Data-structure analysis (or $\DD$-analysis) aims to prove that a suitable
\emph{data-structure integrity constraint} (DSIC) is preserved by the program.
Conversely, a program fails data-structure analysis 
if it is possible to generate a
mal-formed data-structure that violates the DSIC.
More formally, the analysis aims to prove
Hoare triples of the form
$\triple{\DD(\heap, p_1, .., p_n)}{C}{\DD(\heap, q_1, .., q_m)}$
where $C$ is some code fragment (e.g. a function definition),
$\heap$ is the global program heap,
$\{p_1,..,p_n\}$ and $\{q_1,..,q_m\}$ are sets of live pointer variables,
and $\DD$ is a suitable DSIC defined below.
For brevity we abbreviate the DSIC as $\DD$ (without parameters).
If the analysis is successful, then all execution paths through $C$ preserve
$\DD$, and $C$ is said to be $\DD$-safe.

In this section, we formalize the DSIC necessary to implement
our analysis.
Later, we use the formalism as the basis for the implementation 
using the \emph{Satisfiability Modulo Constraint Handling Rules} (SMCHR)
system.

\mysubsubsection{Graph-based Data-structures}
For our purposes, a \emph{data-structure} is a \emph{directed graph} of
\emph{nodes}.
Each node has an associated \emph{type} that corresponds to a
C struct declaration.
A data-structure is considered valid if the following conditions hold,
including:
\begin{enumerate}
\item \label{case:nodes}
    \underline{\emph{Valid nodes}}:
    Each node is a contiguous region of memory whose size is large enough to
    fit the corresponding node type.
    Partially allocated nodes (e.g., size too small) are disallowed.
\item \label{case:ptrs}
    \underline{\emph{Valid pointers}}: All non-null pointers stored within the
    data-structure must point to another valid node.
    Invalid, interior or dangling-pointers are disallowed.
    The null pointer is treated as a special case that indicates the
    non-existence of a link.
\item \label{case:overlap}
    \underline{\emph{Separated nodes}}:
    Nodes must not overlap in memory.
\end{enumerate}
These conditions are desirable for most standard graph-based
data-structures implemented in idiomatic C,
including linked-lists, trees, DAGS, graphs, etc.,
or any other data-structure type that can be described as a graph of nodes
and uses standard pointers.
Our $\DD$-analysis is specific to the above properties, and does
not include any other data-structure property.
In particular, the analysis is \emph{shape neutral}, and does not aim to
analyze for, nor enforce, a given \emph{shape} of the graph.
As such, our $\DD$-analysis is applicable to any graph-based structure,
including
\emph{cyclic} data-structures such as circular linked-lists.

\begin{figure}[t]
\begin{center}
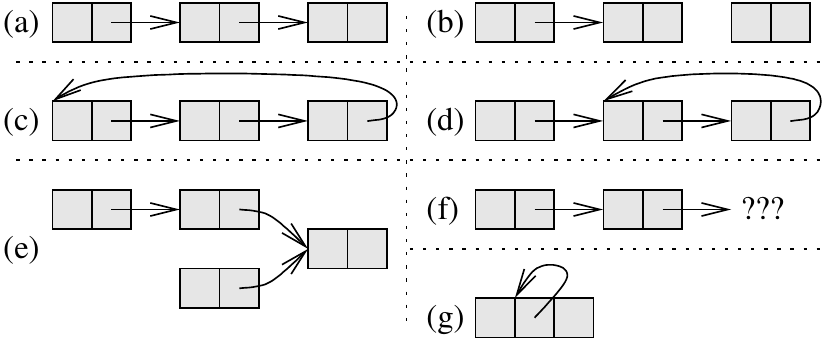
\end{center}
\caption{Various list data-structure shapes.  Here (???) indicates a
dangling pointer.\label{fig:lists}}
\end{figure}

Data-structures in C are declared using some combination
of \texttt{struct} declarations with pointer and data fields.
It is not necessarily apparent what the intended shape of the data-structure
is based on the type declarations alone.
For example, consider the following \texttt{struct} definitions:
\begin{egcode}
    struct list_node {          struct tree_node {
        int val;                    list_node *elem;
        list_node *next; }          tree_node *left; tree_node *right; };
\end{egcode}
For example, the \verb+list_node+ definition can be used to construct: 
(a) a linked-list; (b) a disjoint list; (c) a circular linked-list;
(d) a lasso-list; (e) a list with sharing,
or any combination of the above, 
as illustrated in Figure~\ref{fig:lists}.
Our $\DD$-analysis treats (a), (b), (c), (d), (e)
as graphs of \verb+list_node+s.
List (f) is \emph{invalid} since the last pointer is dangling
(violates \emph{valid pointers}).
Likewise list (g) is invalid since it contains overlapping nodes
(violates \emph{separated nodes}).

The $\DD$-analysis aims to detect code that violates the DSIC.
For example, consider the following ``malicious'' function \verb+make_bad+,
which deliberately constructs a mal-formed linked-list
(with overlapping nodes as per list (g)), and
thus should fail the $\DD$-analysis:
\begin{egcode}
    struct list_node *make_bad(void) {
        struct list_node *xs = malloc(3*sizeof(void *));
        xs->val = 0; xs->next = (struct list_node *)&xs->next;
        xs->next->next = NULL; }
\end{egcode}
Such data-structure violations can lead to counter-intuitive behavior.
For example, consider the following ``benign'' \verb+set+ function that
sets the $n^\mathit{th}$ member of a linked list:
\begin{egcode}
    void set(list_node *xs, int n, int v) {
        while (xs && (n--) > 0) xs = xs->next;
        if (xs) xs->val = v; }
\end{egcode}
Next consider the seemingly benign code
fragment, (\verb+set(xs,1,A); set(xs,1,B);+),
that sets the second node's value to integers \verb+A+ and \verb+B+
respectively.
However, if \verb+xs+ was created with \verb+make_bad+,
the
first call to \verb+set+ clobbers the \verb+next+ field of the first node
with value \verb+A+.
The second node now appears to be at address \verb+A+.
The second call to \verb+set+ executes {\verb+A->val=B+} allowing
for arbitrary memory to be overwritten.

\mysubsubsection{Formalization}
We shall now formalize the integrity constraint $\DD$.
We assume, as given, a set of \emph{node types} 
$\set{Types} = \{\mathit{type}_0, ..., \mathit{type}_n\}$ that are used by the
program, e.g. \verb+list_node+ and \verb+tree_node+ defined above.
We treat each $\mathit{type} \in \set{Types}$ as a set of \emph{fields},
e.g. $\mathtt{tree\_node} = \{\mathtt{elem}, \mathtt{left}, \mathtt{right}\}$.
W.l.o.g., we shall assume all fields are renamed apart.
Given $\set{Types}$,
we define set $\set{Fields}$ as all fields, and
$\set{PtrFields} \subseteq \set{Fields}$
as all fields with a pointer-to-node type.
We also treat $\set{Fields}$ and $\set{PtrFields}$ as sequences by
choosing an arbitrary field ordering.
The sets $\set{Types}$, $\set{Fields}$ and $\set{PtrFields}$ are
derived from all \verb+struct+ declarations in scope.

\begin{figure}
\subfloat[Sub-heap illustrations\label{fig:subheap_1}]{
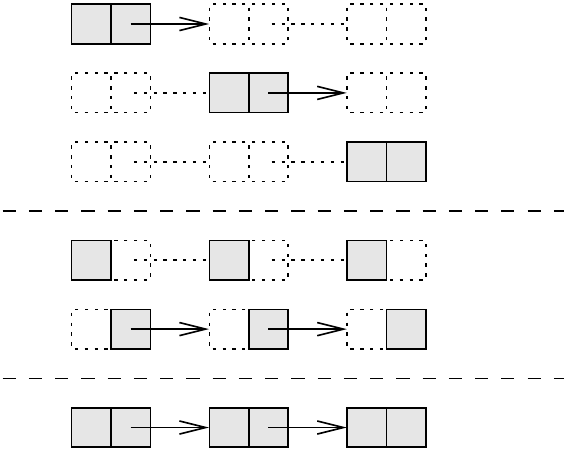
}
~~~~
\vline
\subfloat[Sub-heap expressions\label{fig:subheap_2}]{
\adjustbox{raise=5pc}{
\begin{minipage}{0.5\textwidth}
\begin{align*}
\hone{p}{1} ~\hsep~ \hone{(p{+}1)}{q} & ~~~(N_p) \\
\hone{q}{2} ~\hsep~ \hone{(q{+}1)}{r} & ~~~(N_q) \\
\hone{r}{3} ~\hsep~ \hone{(r{+}1)}{0} & ~~~(N_r) \\
\hone{p}{1} ~\hsep~ \hone{q}{2} ~\hsep~ \hone{r}{3} &
    ~~~(F_\mathtt{val}) \\
\hone{(p{+}1)}{q} ~\hsep~ \hone{(q{+}1)}{r} ~\hsep~ \hone{(r{+}1)}{0}
    & ~~~(F_\mathtt{next})
\end{align*}
\end{minipage}
}
}
\caption{A list example.\label{fig:list-eg}}
\end{figure}

Suppose heap $\heap$ is a valid data-structure,
then $\heap$ is composed of a set of disjoint \emph{node} heaps.
Given a \emph{node pointer} $p$ of type
$(\mathit{T}~*)$, then
a heap $N_p \in \set{Heaps}$ is a \emph{node heap}
for pointer $p$ if it spans the contiguous range of addresses
$p, p+1, .., p + |\mathit{T}|-1$.\footnote{
As a simplification, 
we assume that the $i^{th}$ field is stored in address $p+i$, and
that
$\mathit{sizeof}(\mathtt{int}) = \mathit{sizeof}(\mathtt{void}~*)$.}

An alternative (and unconventional) way to decompose a data-structure is based
on fields.
Given a valid data-structure $\heap$ and a field
$\mathit{field} \in \set{Fields}$, then we define the \emph{field heap}
$F_\mathit{field}$ to be the sub-heap of $\heap$ containing all
address-value pairs associated with the given $\mathit{field}$.
For example, suppose $\heap$ is a 3-node linked-list of type
\verb+list_node+ (defined above), and encodes the sequence $1,2,3$.
We assume the nodes have addresses $p$, $q$, and $r$ respectively.
Heap $\heap$ is therefore representable in Separation
Logic~\cite{reynolds02separation} notation as follows:
\begin{align*}
\hone{p}{1} ~\hsep~ \hone{(p{+}1)}{q} ~\hsep~ \hone{q}{2} ~\hsep~
    \hone{(q{+}1)}{r}
    ~\hsep~ \hone{r}{3} ~\hsep~ \hone{(r{+}1)}{0}
\end{align*}
Heap $\heap$ contains three node sub-heaps $N_p, N_q, N_r \subset \heap$ and
two field sub-heaps $F_\mathtt{val}, F_\mathtt{next} \subset \heap$ defined
in Figure~\ref{fig:subheap_2} and illustrated in Figure~\ref{fig:subheap_1}.
The heap $\heap$ is essentially the disjoint-union of all the field heaps, i.e.,
$\heap \heq F_\mathtt{val} \hsep F_\mathtt{next}$.
Given a set of field heaps, then we can define 
a valid \emph{node-pointer} $p$ as follows:
\begin{definition}[Node Pointers]
\label{def:node}
Let $\mathit{type} \in \set{Types}$ be a node type,
then value $p \in \set{Values}$ is a \emph{$\mathit{type}$-node-pointer} if
\begin{itemize}
    \item[-] $p = 0$ (null pointer) ; or
    \item[-] $p + i \in \dom(F_\mathit{field})$ for each 
          $\mathit{field} \in \mathit{type}$, where
	      $\mathit{field}$ is the $i^{th}$ field of $\mathit{type}$.~$\qed$
\end{itemize}
\end{definition}\noindent
Essentially, a non-null value $p$ is a valid node-pointer for
$\mathit{type} \in \set{Types}$ if the contiguous addresses
$p, p+1, .., p+|\mathit{type}|-1$ are allocated in the corresponding
field heaps.
For example, $q$ from Figure~\ref{fig:subheap_2} is valid since
$q \in \dom(F_\mathtt{val})$ and $q+1 \in \dom(F_\mathtt{next})$.

In order for a data-structure $\heap$ to be valid, all non-null
values $p$ stored in any $\mathit{field} \in \set{PtrField}$ must be
valid node-pointers of the corresponding $\mathit{type}$.
Thus, the graph structure represented by $\heap$ is \emph{closed},
i.e., no invalid (uninitialized, wild, or dangling) links.
\begin{definition}[Closed]
\label{def:closed}
Field heaps $F_{\mathit{field}_1}, .., F_{\mathit{field}_{|\set{Fields}|}}$
are \emph{closed} if
for all $\mathit{field} \in \set{PtrFields}$
and for all $p, v$ such that $\hin{F_\mathit{field}}{p}{v}$,
then $v$ is a valid $T$-node-pointer (Definition~\ref{def:node})
where $\mathit{typeof}(\mathit{field}) = (T~*)$.
$\qed$\end{definition}\noindent
We define:
\begin{itemize}
\item $\mathsf{node}_\mathit{type}(p, F_1, .., F_m)$
to be the
relation satisfying Definition~\ref{def:node} for
field heaps $\{F_1, .., F_m\}$;
\item $\mathsf{closed}(F_1, .., F_m)$
to be the
relation satisfying Definition~\ref{def:closed}.
\end{itemize}
Our DSIC $\DD(\heap)$ is defined as follows:
Given the set $\set{Types}$, we derive the sets $\set{Fields}$ and
$\set{PtrFields}$.
The heap $\heap$ must be partitionable into field heaps, and 
the field heaps must be closed under Definition~\ref{def:closed}:
\begin{align}
    \heap \heq F_1 \hsep \cdots \hsep F_m \land
    \mathsf{closed}(F_1, \cdots, F_m) \tag{\textsc{Closed}}
    \label{eq:closed}
\end{align}
At a given program point, there may be zero or more 
variables $p_1, .., p_n$ pointing to nodes of types
$\mathit{T}_1, .., \mathit{T}_n \in \set{Types}$.
These pointers must be valid under Definition~\ref{def:node}, i.e.:
\begin{align}
\mathsf{node}_{\mathit{T}_1}(p_1, F_1, .., F_m) \land {} \cdots \land
\mathsf{node}_{\mathit{T}_n}(p_n, F_1, .., F_m)
\tag{\textsc{Ptrs}}	\label{eq:pointers}
\end{align}
We can now define the integrity constraint $\DD$:
\begin{definition}[Data-structure Integrity Constraint]
\label{def:dsi}
The \emph{data-structure integrity constraint} $\DD$
is defined by combining the above
components (via textual substitution) as follows:
\begin{align*}
\DD(\heap, p_1, .., p_n) ~\eqdef~
	\exists F_1, \cdots, F_m:
	\text{\rm (\ref{eq:closed})} \land \text{\rm (\ref{eq:pointers})}
    ~~~~\qed
\end{align*}
\end{definition}

%%%%%%%%%%%%%%%%%%%%%%%%%%%%%%%%%%%%%%%%%%%%%%%%%%%%%%%%%%%%%%%%%%%%%%%%%%%%%
\mysubsubsection{Spatial Memory Safety}
The basic analysis of Section~\ref{sec:closed} assumes that all
memory outside the data-structure is \emph{unmapped},
which is unrealistic in practice.
We extend our memory model to account for
some arbitrary context of mapped memory by
splitting the global heap $\heap$ into 
a \emph{footprint heap} $\foot$ and
a \emph{context heap} $\fram$ as follows:
$\heap \heq \foot ~\hsep~ \fram$.
The data-structure resides in the footprint heap $\foot$, and the
context heap $\fram$ represents any other mapped memory, such as
the stack, globals, free-lists, etc.
Buggy code may access $\fram$ via a \emph{spatial memory error}, such as an
object bounds overflow, thus violating memory safety.
To detect such errors,
we extend the DSIC as follows:
\begin{definition}[Data-structure Integrity Constraint \RN{2}]
\label{def:dsi2}
Let $\DD$ be the basic data-structure integrity constraint from
Definition~\ref{def:dsi}, then:
\begin{align*}
    \DD_M(\heap, \fram, p_1, .., p_n) ~\eqdef~
        \exists \foot : \heap \heq \foot \hsep \fram \land 
        \DD(\foot, p_1, .., p_n)
        ~~~~ \qed
\end{align*}
\end{definition}\noindent
Finally, to prove spatial memory safety, it must be shown that
\begin{align}\label{eq:sms}
    \triple{\DD_M(\heap, \fram)}{P; (x := *p)}{p \not\in \dom(\fram)}
        \tag{\textsc{Spatial Memory Safety}}
\end{align}
for all reads of pointer $p$.
Likewise, we similarly must verify all writes $(*p := x)$.

%%%%%%%%%%%%%%%%%%%%%%%%%%%%%%%%%%%%%%%%%%%%%%%%%%%%%%%%%%%%%%%%%%%%%%%%%%%%%
\section{Solving for Data-Structures}\label{sec:solver}

The $\DD_M$-analysis depends on determining the validity of the
\emph{Verification Conditions} (VCs) generated by
symbolic execution, which are of the form:
\begin{align}
    \label{eq:vc0} \tag{\textsc{VC}}
    \begin{array}{c}
    \mathit{path}(\heap, \fram, F_1, .., F_m) 
    \models 
    \exists \foot', F'_1, .., F'_m :
    \mathit{post}(\heap, \fram, F'_1, .., F'_m)
    \end{array}
\end{align}
where $\mathit{path}$ is the path constraint, 
$\mathit{post}$ is the post-condition,
$\heap$ is the global heap, $\fram$ is the context heap,
$F_1, .., F_m$ are the initial field heaps,
$\foot'$ is the modified footprint and
$F'_1, .., F'_m$ are the modified field heaps.
Validity can be established by a two-step process:
(1) generating witnesses $W$ for $\foot', F'_1, .., F'_m$; and
(2) proving that $(W \land \mathit{path} \land \neg \mathit{post})$ is
\emph{unsatisfiable} using a solver.
Witnesses $W$ are built using
the following schema:
\begin{align*}
\begin{array}{ll}
    W  \eqdef (\foot' = \heap - \fram) \land X_{f \in \set{Fields}} \land
        Y_{p \in \set{Allocs}} &
    Y_p \eqdef Z^p_{f \in \mathsf{typeof}(*p)} \\
    X_f \eqdef
    F'_\mathit{f} \subseteq \foot' \land
    \dom(F_\mathit{f}) \subseteq \dom(F'_\mathit{f})  &
    Z^p_f \eqdef
    p + \mathsf{offsetof}(\mathsf{typeof}(*p), f) \in \dom(F'_\mathit{f})
\end{array}
\end{align*}
Here $T_{x \in \{a, .., z\}}$ is shorthand for $(T_a \land .. \land T_z)$,
and $\set{Allocs}$ is defined to be all allocated pointers
(i.e. $p = \mathsf{malloc}(..))$)
in $\mathit{path}$.
Intuitively, $\foot'$ is the heap difference $\heap - \fram$,
and $F'_\mathit{field}$ is the heap that is
(1) a sub-heap of $\foot'$, and
(2) has the same domain as $F_\mathit{field}$ save for any new
addresses created by allocations.

The next step is to prove that the quantifier-free formula
$(W \land \mathit{path} \land \neg \mathit{post})$ is unsatisfiable.
For this we use a combination of an integer solver,
an extension of the heap solver from~\cite{duck13heaps} (a.k.a., the
$\HH$-solver) and a specialized solver
for data-structure constraints defined below (a.k.a., the $\DD$-solver).
The $\DD$-solver is implemented using the
\emph{Constraint Handling Rules} (CHR) solver
language~\cite{fruhwirth09chr} using the following basic solver schema
customized for the types declared by the program:
\begin{align*}
\mathsf{closed}(F_0, .., F_m) \land \hin{F_\mathit{f}}{p}{v} 
\prop & ~
\mathsf{node}_\mathit{type}(v, F_0, .., F_m) & \hspace{-4cm} \text{where}~\mathit{f}\in\set{PtrFields}
    ~\text{and}~ \mathit{f}\in\mathit{type} \\
\mathsf{node}_\mathit{type}(p, F_0, .., F_m) 
\prop & ~
p = 0 \lor \big (\bigwedge_{f \in \mathit{type}} 
p + \mathsf{offsetof}(\mathit{type}, f) \in \dom(F_f)
\big )
\end{align*}
The rules encode the greatest relations satisfying
Definitions~\ref{def:closed} and~\ref{def:node} respectively.

Given a set of types, the schema is automatically instantiated to generate
a specialized CHR solver (the $\DD$-solver) for the corresponding
$\DD$-constraints.
The $\DD$-solver can then be used to solve VCs using the
\emph{Satisfiability Modulo Constraint Handling Rules} (SMCHR)
system~\cite{duck12smchr,duck13smchr}
in combination with existing heap, integer, and equality built-in
solvers.
For example, assuming $\set{Types} = \{\mathtt{list\_node}\}$, the
corresponding $\DD$-solver is:
\begin{align*}
& \mathsf{closed}(F_\mathtt{val}, F_\mathtt{next}) \land
    \hin{F_\mathtt{next}}{p}{v} \prop 
    \mathsf{node}(v, F_\mathtt{val}, F_\mathtt{next}) \\
& \mathsf{node}(p, F_\mathtt{val}, F_\mathtt{next}) \prop 
p = 0 \lor \big (p \in \dom(F_\mathtt{val}) \land
         p {+} 1 \in \dom(F_\mathtt{next}) \big)
\end{align*}
Consider the statement $S$ $=$ (\verb+xs=xs->next+).
Assuming that the $\DD_M$ property (Definition~\ref{def:dsi2})
holds before $S$, we can prove $S$
to be memory safe using the VC:
\begin{align*}
\heap \heq \foot \hsep \fram \land 
\foot \heq F_\mathtt{val} \hsep F_\mathtt{next} \land {} 
    \mathsf{node}(\mathit{xs}, F_\mathtt{val}, F_\mathtt{next})
    \models
    \mathit{xs}{+}1 \not\in \dom(\fram)
\end{align*}
This VC is valid iff the constraints in Figure~\ref{fig:solve_vc} 1) are
unsatisfiable.
The solver steps are shown in Figure~\ref{fig:solve_vc}.
Here ($\HH$), ($\DD$), and (I) represent inferences made by the
$\HH$-solver,
$\DD$-solver, and integer solver respectively.
The constraints used by each inference step are \underline{underlined}.
Step 2) introduces a disjunction which leads to two branches 3a) and 3b).
Since all branches lead to $\false$ the original goal is unsatisfiable,
hence proving the VC is valid.

One of the main features of the SMCHR system is the ability to extend
existing built-in solvers with new constraints implemented using CHR.
For example, the built-in $\HH$-solver works by propagating
\emph{heap element} constraints of the form $(\hin{H}{p}{v})$ or
$(p \in \dom(H))$~\cite{duck13heaps}.
The $\DD$-solver extends the $\HH$-solver with new rules
that interact with these constraints.
Solver communication is two-way, e.g., the $\DD$-solver may propagate
new element constraints (e.g., the $\mathtt{node}$ rule),
or may match element constraints
propagated by the $\HH$-solver (e.g., the $\mathtt{closed}$ rule). 

\begin{figure*}[t]
\centering
{\small
\begin{align*}
    \begin{array}{rll}
    1) & \{\heap \heq \foot \hsep \fram,
           \foot \heq F_\mathtt{val} \hsep F_\mathtt{next},
           \mathsf{node}(\mathit{xs}, F_\mathtt{val}, F_\mathtt{next}),
        \underline{\mathit{xs}{+}1 \in \dom(\fram)}\} & (\HH) \\
    2) & \{\heap \heq \foot \hsep \fram,
           \foot \heq F_\mathtt{val} \hsep F_\mathtt{next},
        \underline{\mathsf{node}(\mathit{xs}, F_\mathtt{val}, F_\mathtt{next})},
        \mathit{xs}{+}1 \in \dom(\fram), \mathit{xs}>0\} & (\DD) \\
    \noalign{\vskip 0.5em} 
    3a) & \{\heap \heq \foot \hsep \fram,
        \foot \heq F_\mathtt{val} \hsep F_\mathtt{next},
        \mathsf{node}(\mathit{xs}, F_\mathtt{val}, F_\mathtt{next})),
        \mathit{xs}{+}1 \in \dom(\fram), \underline{\mathit{xs}>0},
        \underline{\mathit{xs} = 0} \} & (I) \\
    4a) & \false \\
    \noalign{\vskip 0.5em} 
    3b) & 
        \begin{array}{l}
        \{
        \underline{\heap \heq \foot \hsep \fram},
        \underline{\foot \heq F_\mathtt{val} \hsep F_\mathtt{next}},
        \mathsf{node}(\mathit{xs}, F_\mathtt{val}, F_\mathtt{next}), 
        \underline{\mathit{xs}{+}1 \in \dom(\fram)}, \mathit{xs}>0, \\
        ~~\underline{\mathit{xs}{+}1 \in \dom(F_\mathtt{next})}
        \}
        \end{array}
        & (\HH) \\
    4b) & \false
    \end{array}
\end{align*}
}
\caption{Solver steps for an example memory safety VC.\label{fig:solve_vc}}
\end{figure*}

It is also possible to instantiate the $\DD$-solver schema with
multiple types.
For example, assuming
$\set{Types} = \{\mathtt{list\_node}, \mathtt{tree\_node}\}$,
the following $\DD$-solver rules will be generated:
\begin{align*}
& \mathsf{closed}(F_\mathtt{val}, F_\mathtt{next}, F_\mathtt{elem},
    F_\mathtt{left}, F_\mathtt{right}) \land
    \hin{F_\mathtt{next}}{p}{v} \prop 
    \mathsf{node}_\mathtt{list\_node}(v, F_\mathtt{val}, F_\mathtt{next}) \\
& \mathsf{closed}(F_\mathtt{val}, F_\mathtt{next}, F_\mathtt{elem},
    F_\mathtt{left}, F_\mathtt{right}) \land
    \hin{F_\mathtt{left}}{p}{v} \prop 
    \mathsf{node}_\mathtt{tree\_node}(v, F_\mathtt{elem}, F_\mathtt{left},
        F_\mathtt{right}) \\
& \mathsf{closed}(F_\mathtt{val}, F_\mathtt{next}, F_\mathtt{elem},
    F_\mathtt{left}, F_\mathtt{right}) \land
    \hin{F_\mathtt{right}}{p}{v} \prop 
    \mathsf{node}_\mathtt{tree\_node}(v, F_\mathtt{elem}, F_\mathtt{left},
        F_\mathtt{right}) \\
& \mathsf{node}_\mathtt{list\_node}(p, F_\mathtt{val}, F_\mathtt{next}) \prop 
p = 0 \lor \big (p \in \dom(F_\mathtt{val}) \land
         p {+} 1 \in \dom(F_\mathtt{next}) \big) \\
& \mathsf{node}_\mathtt{tree\_node}(p, F_\mathtt{elem}, F_\mathtt{left},
    F_\mathtt{right}) \prop \\
& ~~~~ ~~~~ ~~~~ ~~~~
p = 0 \lor \big (p \in \dom(F_\mathtt{elem}) \land
         p {+} 1 \in \dom(F_\mathtt{left}) \land
         p {+} 2 \in \dom(F_\mathtt{right}) \big)
\end{align*}

\mysubsubsection{Handling Negation}
Some VCs may contain negated $\DD$-constraints
$\mathsf{node}$ and $\mathsf{closed}$.
We can eliminate all negated $\DD$-constraints by applying the
following rewrite rules:
\begin{align*}
\begin{array}{rcl}
 \begin{array}{c}
 \neg \mathsf{closed}(F_1, .., F_m) \\ \longrightarrow \\
     \bigvee\limits_{
         \mathit{field} \in \set{PtrFields}
         } 
     \left (
     \begin{array}{c}
     \hin{F_\mathit{field}}{s}{t} \land {} \\
     \neg \mathsf{node}_\mathit{type}(t, F_1, .., F_m)
     \end{array}
     \right )
 \end{array} & ~~~~ &
 \begin{array}{c}
 \neg \mathsf{node}_\mathit{type}(p, F_1, .., F_m) \\ \longrightarrow \\
    p \neq 0 \land \big (
    \bigvee\limits_{
        \mathit{field} \in \mathit{type}
        }
        p + i \not\in \dom(F_\mathit{field})
    \big ) \\
 \end{array}
 \end{array}
\end{align*}
where $\mathit{typeof}(\mathit{field}) = (\mathit{type}~*)$, index
$i = \mathit{offsetof}(\mathit{field}, \mathit{type})$,
and variables $s, t$ are assumed fresh.
These rules 
implement the negations of Definitions~\ref{def:node} and~\ref{def:closed}
respectively.
For example, $\neg \mathsf{closed}(F_\mathtt{val}, F_\mathtt{next})$
can be rewritten to
\begin{align*}
\hin{F_\mathtt{next}}{s}{t} \land t \neq 0 \land
    \big (
        t \not\in \dom(F_\mathtt{val}) \lor
        t{+}1 \not\in \dom(F_\mathtt{next})
    \big )
\end{align*}
That is, in order for $\mathsf{closed}(F_\mathtt{val}, F_\mathtt{next})$
to be violated, there must exist a heap cell $\hin{F_\mathtt{next}}{s}{t}$
such that (1) $t$ is non-null, and (2) $t$ does not point to a valid
node, i.e. $(t \not\in \dom(F_\mathtt{val}))$ or
$(t{+}1 \not\in \dom(F_\mathtt{next}))$.
As with the CHR rules, the rewrite rules are generated automatically.

%%%%%%%%%%%%%%%%%%%%%%%%%%%%%%%%%%%%%%%%%%%%%%%%%%%%%%%%%%%%%%%%%%%%%%%%%%%%%%
\section{Experiments}\label{sec:experiments}

We have implemented a prototype $\DD_M$-analysis tool (called $\DD$-tool)
as a~\cite{llvm} plug-in.
The tool takes as input a C program that is first
converted into the LLVM \emph{Intermediate Representation} (IR) 
using the \texttt{clang} front-end.
The plug-in 
implements the $\DD_M$-analysis as
described in Section~\ref{sec:closed}, and
automatically generates a specialized solver
as described in Section~\ref{sec:solver}.
The VCs are solved using a constraint solver back-end, namely the
\emph{Satisfiability Modulo Constraint Handling Rules} 
(SMCHR)~\cite{duck12smchr} system, using the generated
solver in combination with existing built-in heap, integer and equality
solvers.
The SMCHR system also supports goal transformation using rewrite rules,
which is used to implement negation.
The result is either SAFE if all generated VCs are proved valid, or
(possibly) UNSAFE otherwise.
The entire process (i.e. compilation, VC generation,
solver generation, and solving) is automatic.
All experiments were run on an
Intel i7-4770 CPU clocked at 3.4GHz.

\newcommand{\flistwidth}{4.5cm}
\newcommand{\structwidth}{1.5cm}
\begin{figure*}[t]
{ \small
\begin{center}
\begin{oldtabular}{|c|c|c|c|l|r||r|r|r||r|r|r||}
\cline{1-12}
 & & & & & & \multicolumn{3}{c||}{\texttt{malloc}} & 
             \multicolumn{3}{c||}{\texttt{zmalloc}} \\
\cline{1-12}
~ & \emph{Struct.} & \textsf{\#Nd} & \textsf{Sh?} & \emph{Func.} &
    \textsf{LOC} & \textsf{Time} & \textsf{$\DD$} & \textsf{M} &
    \textsf{Time} & \textsf{$\DD$} & \textsf{M} \\
\cline{1-12}
\parbox[t]{2mm}{\multirow{3}{*}{\rotatebox[origin=c]{90}{GLib~~~~~ }}} 
&
    \begin{minipage}{\structwidth}
    \begin{center}
    \emph{doubly linked-list}
    \end{center}
    \end{minipage}
 & 1+3 & \xmark &
     \begin{minipage}{\flistwidth}
     \flushleft \tiny
     \vspace{0.3em}
     \verb+foreach+
     \verb+append+
     \verb+last+
     \verb+prepend+
     \verb+insert+
     \verb+nth+
     \verb+insert_before+
     \verb+concat+
     \verb+remove+
     \verb+_remove_link+
     \verb+remove_all+
     \verb+remove_link+
     \verb+delete_link+
     \verb+copy+
     \verb+copy_deep+
     \verb+reverse+
     \verb+nth_prev+
     \verb+nth_data+
     \verb+find+
     \verb+find_custom+
     \verb+position+
     \verb+index+
     \verb+first+
     \verb+length+
     \verb+insert_sorted+
     \verb+insert_sorted_real+
     \verb+insert_sorted_with_data+
     \verb+sort+
     \verb+sort_real+
     \verb+sort_with_data+
     \verb+sort_merge+
     \vspace{0.3em}
     \end{minipage}
    & 545 & 0.52 & {28} & \textbf{33} & 0.48 & \textbf{33} & \textbf{33} \\
\cline{2-12}
&
    \begin{minipage}{\structwidth}
    \begin{center}
    \emph{singly linked-list}
    \end{center}
    \end{minipage}
& 1+2 & \xmark &
    \begin{minipage}{\flistwidth}
    \flushleft \tiny
    \vspace{0.3em}
    \verb+foreach+
    \verb+append+
    \verb+last+
    \verb+prepend+
    \verb+insert+
    \verb+insert_before+
    \verb+concat+
    \verb+remove+
    \verb+remove_all+
    \verb+remove_link+
    \verb+_remove_link+
    \verb+delete_link+
    \verb+copy+
    \verb+copy_deep+
    \verb+reverse+
    \verb+nth+
    \verb+nth_data+
    \verb+find+
    \verb+find_custom+
    \verb+position+
    \verb+index+
    \verb+length+
    \verb+insert_sorted+
    \verb+insert_sorted_real+
    \verb+insert_sorted_with_data+
    \verb+sort+
    \verb+sort_real+
    \verb+sort_with_data+
    \verb+sort_merge+
    \vspace{0.3em}
    \end{minipage}
    & 507 & 0.25 & {27} & \textbf{31} & 0.26 & \textbf{31} & \textbf{31} \\
\cline{2-12}
&
    \begin{minipage}{\structwidth}
    \begin{center}
    \emph{red-black tree}
    \end{center}
    \end{minipage}
& 2+14 & \xmark &
     \begin{minipage}{\flistwidth}
     \flushleft \tiny
     \vspace{0.3em}
     \verb+remove_all+
     \verb+insert+
     \verb+replace+
     \verb+remove+
     \verb+steal+
     \verb+lookup+
     \verb+find_node+
     \verb+foreach+
     \verb+first_node+
     \verb+node_next+
     \verb+traverse+
     \verb+node_pre_order+
     \verb+node_in_order+
     \verb+node_post_order+
     \verb+search+
     \verb+node_search+
     \verb+height+
     \verb+nnodes+
     \vspace{0.3em}
     \end{minipage}
    & 858 & 4.05 & \textbf{27} & \textbf{27} & 4.02 & \textbf{27} & \textbf{27} \\
\cline{1-12}
\parbox[t]{2mm}{\multirow{2}{*}{\rotatebox[origin=c]{90}{VF}}}
&
    \begin{minipage}{\structwidth}
    \begin{center}
        \emph{binary tree}
    \end{center}
    \end{minipage}
& 1+3 & \xmark &
    \begin{minipage}{\flistwidth}
    \flushleft \tiny
    \vspace{0.3em}
\verb+init_tree+ \verb+free_tree+ \verb+contains+ \verb+add+ \verb+maximum+
\verb+remove+ \verb+main+
    \vspace{0.3em}
    \end{minipage}
    & 157 & 0.20 & {6} & \textbf{7} & 0.20 & \textbf{7} & \textbf{7} \\
\cline{2-12}
&
    \begin{minipage}{\structwidth}
    \begin{center}
        \emph{graph}
    \end{center}
    \end{minipage}
& 1+4 & \cmark &
    \begin{minipage}{4cm}
    \vspace{0.5em}
    {\tiny \verb+schorr_waite+}
    \vspace{0.5em}
    \end{minipage}
    & 38 & 0.06 & \textbf{1} & \textbf{1} & 0.06 & \textbf{1} & \textbf{1} \\
\cline{1-12}
\parbox[t]{2mm}{\multirow{2}{*}{\rotatebox[origin=c]{90}{~libf~~~~}}}
&
    \begin{minipage}{\structwidth}
    \begin{center}
        \emph{234 tree}
    \end{center}
    \end{minipage}
& 1+8 & \cmark &
    \begin{minipage}{\flistwidth}
    \flushleft \tiny
    \vspace{0.3em}
\verb+_tree_singleton+ 
\verb+tree_is_empty+
\verb+tree_is_singleton+
\verb+_tree_search+ 
\verb+tree_search_any+ 
\verb+tree_search_min+ 
\verb+tree_search_max+ 
\verb+tree_search_lt+ 
\verb+_tree_size+ 
\verb+tree_depth+ 
\verb+_tree_foldl+
\verb+_tree_map+
    \vspace{0.3em}
    \end{minipage}
    & 452 & 5.94 & \textbf{12} & \textbf{12} & 5.98 & \textbf{12} & \textbf{12} \\
\cline{2-12}
&
    \begin{minipage}{\structwidth}
    \begin{center}
        \emph{23 finger tree}
    \end{center}
    \end{minipage}
& 3+17 & \cmark &
    \begin{minipage}{\flistwidth}
    \flushleft \tiny
    \vspace{0.3em}
\verb+_seq_is_empty+
\verb+_seq_length+
\verb+dig_length+
\verb+tree_length+
\verb+_seq_lookup+
\verb+tree_lookup+
\verb+dig_lookup+
\verb+seq_push_front+
\verb+_seq_replace_front+
\verb+_seq_peek_front+
\verb+_seq_foldl+
\verb+tree_foldl+
\verb+dig_foldl+
\verb+_seq_map+
\verb+tree_map+
\verb+dig_map+
    \vspace{0.3em}
    \end{minipage}
    & 830 & 62.93 & \textbf{21} & \textbf{21} & 64.09 & \textbf{21} & \textbf{21} \\
\cline{1-12}
\end{oldtabular}
\end{center}
}
\caption{$\DD_M$-analysis benchmarks for safe library code.\label{fig:safety}}
\end{figure*}

\mysubsubsection{$\DD$-tool Verification on Safe Modules}
As the $\DD$-tool can analyze partial programs---one important
use case is to analyze libraries (or modules).
Figure~\ref{fig:safety} tests the $\DD$-tool against several
memory safe functions that manipulate data-structures sourced from 
the following C libraries:
the GNU GLib library (version 2.38.0) representative of real library code used
by a large number of programs,
Verifast (abbr. VF)~\cite{jacobs11verifast} distribution (manually verified
safe modules),
and the \verb+libf+ library\footnote{\url{https://github.com/GJDuck/libf}}.
These benchmarks test
a wide variety of data-structure types and shapes, including:
singly-linked-lists,
doubly-linked-lists,
red-black-trees,
binary-trees,
binary-graphs, 
234-trees, and
23-finger-trees~\cite{hinze06finger}.
In Figure~\ref{fig:safety},
\textsf{\#Nd} is the pair $\mathit{nodes}{+}\mathit{fields}$ where
$\mathit{nodes}$ is the number of node types and $\mathit{fields}$ is
the number of fields used by the data-structure,
(\textsf{Sh?}) indicates whether the data-structure is designed for
\emph{sharing}
(i.e., each node may have multiple parent nodes),
\textsf{LOC} is the total source-lines-of-code, \textsf{Time} is the
total time (in seconds),
and $\DD$/\textsf{M} is the number of functions proven $\DD$-safe/memory-safe
(see~(\ref{eq:sms})) respectively.
Ideal results are highlighted in \textbf{bold}.
We test two versions of the analysis: \emph{uninitializing} \texttt{malloc}
and \emph{zero-initializing} \texttt{zmalloc}, such as that used
by~\cite{boehm88garbage}.

The GLib benchmarks represent standard C data-structures, namely
linked-lists (singly or doubly) and trees (red-black balanced binary trees).
The typical usage of GLib assumes no data-structure sharing, so each node
has at most one parent node.
The Verifast benchmarks contain an alternative tree implementation, and
binary graphs.
For the graph benchmark, we verify the $\DD_M$-safety of the
\emph{Schorr Waite algorithm}.\footnote{
Verifast verifies the Schorr Waite algorithm for trees
and not general graphs.
}
The \verb+libf+ library implements 234-trees (for immutable maps and sets)
and 23-finger-trees~\cite{hinze06finger} (for immutable sequences).
Finger trees are a relatively complex data-structure, with 3 node types and
an intricate shape.
Furthermore, the \verb+libf+ library employs automatic memory management
via garbage collection, and is specifically designed to allow
data-structure sharing.

Figure~\ref{fig:safety} shows that the $\DD_M$-analysis  
performs well on library data structures, with
all functions automatically verified to be safe under $\mathsf{zmalloc}$.
This result is sufficient for programs using allocators such
as~\cite{boehm88garbage}.
For $\mathsf{malloc}$ 
the results were less precise, with some $\DD$ VCs failing because of
partially initialized data-structures.
As future work, the $\DD$-analysis could be improved
by considering weaker forms of the DSIC to account for uninitialized fields.

\begin{figure*}
{\small
\begin{center}
\begin{oldtabular}{|c|c||c|c|c|c|c|c|}
\cline{1-8}
\emph{Type} & \emph{Tool} & \textsf{Lang.} & \textsf{Static?} &
    \textsf{Auto?} & \textsf{Modular?} & \textsf{Mem. Safety?} &
    \textsf{Data. Structs?} \\
\cline{1-8}
\parbox[t]{2mm}{\multirow{5}{*}{\rotatebox[origin=c]{90}{Sep. Log.}}}
& \textsc{SmallFoot}    & custom     & \cmark & \xmark & \cmark & \cmark &
    \emph{lists, trees} \\
& \textsc{SpaceInvader} & C & \cmark & \cmark & \xmark & \cmark &
    \emph{lists} \\
& \textsc{SLAyer}       & C & \cmark & \cmark & \xmark & \cmark &
    \emph{lists} \\
& Predator              & C & \cmark & \cmark & \xmark & \cmark &
    \emph{lists} \\
& VeriFast              & C & \cmark & \xmark & \cmark & \cmark &
    \emph{any} (limited) \\
\cline{1-8}
\parbox[t]{2mm}{\multirow{2}{*}{\rotatebox[origin=c]{90}{BMC}}}
& LLBMC     & C & \cmark & \cmark & \xmark & \emph{limited} & 
    \emph{any} (bounded) \\
& CBMC      & C & \cmark & \cmark & \xmark & \emph{limited} & 
    \emph{any} (bounded) \\
\cline{1-8}
BC & LowFat & C/C++ & \xmark & \cmark & \xmark &
    \emph{limited} & \emph{any} \\
\cline{1-8}
& $\DD$-tool & C & \cmark & \cmark & \cmark & \emph{limited} &
    \emph{any} \\
\cline{1-8}
\end{oldtabular}
\end{center}
}
\caption{Summary of related tools and trade-offs.\label{fig:toolcomp}}
\end{figure*}

%%%%%%%%%%%%%%%%%%%%%%%%%%%%%%%%
\mysubsubsection{Comparing Memory Safety Tools}
We also compare against several existing memory safety analysis tools,
summarized in Figure~\ref{fig:toolcomp}, which are classified into
four main types:
(1) our $\DD_M$-analysis tool;
(2) \emph{Separation Logic}-based analysis tools such as
    \textsc{SmallFoot}~\cite{berdine05smallfoot},
    \textsc{SpaceInvader}~\cite{disefano06local},
    SLAyer~\cite{berdine11slayer},
    Predator~\cite{dudka11predator,dudka13byte} and 
    Verifast~\cite{jacobs11verifast};
(3) \emph{Bounded Model Checking} (BMC) based analysis tools such as
    LLBMC~\cite{merz12llbmc} and CBMC~\cite{kroening14cbmc}; and
(4) \emph{Bounds Checking} (BC) instrumentation tools such as
    LowFat~\cite{duck16heap,duck17stack}.
Different approaches 
have different trade-offs:
(\textsf{Static?}) whether the tool is based on static program analysis;
(\textsf{Auto?}) whether the tool is fully automatic, or requires
    user intervention (e.g., annotations);
(\textsf{Modular?}) whether the tool can be used to analyze
    individual functions (i.e., suitable for libraries), or requires
    a complete program including an entry point
    (e.g., the \verb+main+ function);
(\textsf{Mem. Safety?}) whether the tool checks for all types
    of classical memory errors (including null-pointers),
    otherwise the tool is
    \emph{limited} to some specific subset;
(\textsf{Data. Structs?}) lists the types of graph-based data-structure
    that are compatible with the tool.
Clearly there are different trade-offs between the different classes of
tools.
Separation Logic-based tools can be used to prove ``full'' memory safety,
including null-pointer and temporal memory errors (use-after-free),
but are either
(1) limited to narrow classes of data-structures, such as lists, or
(2) are not automatic and require user annotations.
In contrast, the $\DD$-tool targets specific memory errors (spatial),
but is not limited to specific types of data-structures.
The $\DD$-tool does not target use-after-free errors which
typically depend on the data-structure shape---i.e.,
that the freed node is not \emph{shared} thereby creating a dangling pointer.
Only 3 tools (\textsc{SmallFoot}, Verifast, $\DD$-tool) are modular.
In contrast,
the other tools require the whole program for analysis.
BMC-based tools are automatic but check for weaker notions of memory safety.
Bounded model checking may also fail to detect errors that are
beyond the search horizon of the tool.
Dynamic bounds checking differs from static analysis-based methods in that
it cannot be used to prove that the whole program is error free.
At best, dynamic analysis tools can only prove that specific paths
are error free.

\begin{figure*}[t]
\begin{center}
{ \small
\begin{oldtabular}{|l|c||c|c|c|c|c|c||c|c|c|c|c||c|c|c|c|c|}
\cline{1-18}
& & \multicolumn{6}{c||}{List} & \multicolumn{5}{c||}{DAG} &
    \multicolumn{5}{c|}{Graph} \\
\cline{1-18}
\emph{Type} & &
    \multicolumn{3}{c|}{Analysis} &
    \multicolumn{2}{c|}{BMC} &
    BC &
    \multicolumn{3}{c|}{Analysis} &
    \multicolumn{2}{c||}{BMC} &
    \multicolumn{3}{c|}{Analysis} &
    \multicolumn{2}{c|}{BMC} \\
\cline{1-18}
\emph{Prog.} & \textsf{Safe?} &
    \rotatebox[origin=c]{90}{\scriptsize $\DD$-tool} &
    \rotatebox[origin=c]{90}{\scriptsize Invader} &
    \rotatebox[origin=c]{90}{\scriptsize Predator} &
    \rotatebox[origin=c]{90}{\scriptsize CBMC} &
    \rotatebox[origin=c]{90}{\scriptsize LLBMC} &
    \rotatebox[origin=c]{90}{\scriptsize LowFat} &
    %%%%
    \rotatebox[origin=c]{90}{\scriptsize $\DD$-tool} &
    \rotatebox[origin=c]{90}{\scriptsize Invader} &
    \rotatebox[origin=c]{90}{\scriptsize Predator} &
    \rotatebox[origin=c]{90}{\scriptsize CBMC} &
    \rotatebox[origin=c]{90}{\scriptsize LLBMC} &
    %%%%
    \rotatebox[origin=c]{90}{\scriptsize $\DD$-tool} &
    \rotatebox[origin=c]{90}{\scriptsize Invader} &
    \rotatebox[origin=c]{90}{\scriptsize Predator} &
    \rotatebox[origin=c]{90}{\scriptsize CBMC} &
    \rotatebox[origin=c]{90}{\scriptsize LLBMC} \\
\cline{1-18}
{\small \verb+overlap_node+}   & \multirow{8}{*}{N} &
    \textbf{U} & c & \textbf{U} & {n} & {n} & n &
    \textbf{U} & c & t.o. & {b} & \textbf{B} & 
    \textbf{U} & c & \textbf{U} & \textbf{B} & \textbf{B} \\
{\small \verb+wrong_node+}    & &
    \textbf{U} & \textbf{U} & \textbf{U} & {n} & \textbf{B} & \textbf{C} &
    \textbf{U} & {s} & t.o. & \textbf{B} & b.r. & 
    \textbf{U} & t.o. & t.o. & \textbf{B} & \textbf{B} \\
{\small \verb+wrong_size+}    & &
    \textbf{U} & {s} & \textbf{U} & {n} & \textbf{B} & \textbf{B} &
    \textbf{U} & {s} & \textbf{U} & \textbf{B} & \textbf{B} & 
    \textbf{U} & t.o. & \textbf{U} & \textbf{B} & \textbf{B} \\
{\small \verb+not_array+}     & &
    \textbf{U} & {s} & \textbf{U} & \textbf{B} & \textbf{B} &
    \textbf{B} &
    \textbf{U} & {s} & t.o. & \textbf{B} & \textbf{B} & 
    \textbf{U} & t.o. & t.o. & \textbf{B} & \textbf{B} \\
{\small \verb+cast_int+}      & &
    \textbf{U} & \textbf{U} & \textbf{U} & \textbf{B} & \textbf{B} & n &
    \textbf{U} & \textbf{U} &t.o. & \textbf{B} & \textbf{B} &
    \textbf{U} & t.o. & t.o. & \textbf{B} & b.r. \\
{\small \verb+uninit_ptr+}    & &
    \textbf{U} & c & \textbf{U} & \textbf{B} & \textbf{B} & n &
    \textbf{U} & \textbf{U} & t.o. & \textbf{B} & \textbf{B} &
    \textbf{U} & t.o. & \textbf{U} & \textbf{B} & \textbf{B} \\
{\small \verb+uninit_ptr_stk+} & &
    \textbf{U} & \textbf{U} & \textbf{U} & \textbf{B} & \textbf{B} & n &
    \textbf{U} & \textbf{U} & t.o. & \textbf{B} & \textbf{B} &
    \textbf{U} & t.o. & t.o. & \textbf{B} & \textbf{B} \\
{\small \verb+arith_ptr+}     & &
    \textbf{U} & {s} & \textbf{U} & \textbf{B} & \textbf{B} & \textbf{B} &
    \textbf{U} & {s} & t.o. & \textbf{B} & \textbf{B} & 
    \textbf{U} & t.o. & t.o. & \textbf{B} & \textbf{B} \\
\cline{1-18}
\small{\verb+safe+}           & Y &
    \textbf{S} & \textbf{S} & \textbf{S} & \textbf{N} & \textbf{N} & \textbf{N} &
    \textbf{S} & \textbf{S} & t.o. & b.r. & b.r. & 
    \textbf{S} & t.o. & t.o. & b.r. & b.r. \\
\cline{1-18}
\emph{Score}
    & &
    \textbf{9} & 4 & \textbf{9} & 6 & 8 & 5 &
    \textbf{9} & 4 & 1 & 7 & 7 &
    \textbf{9} & 0 & 3 & 8 & 7 \\
\cline{1-18}
\end{oldtabular}
}
\end{center}
\caption{Comparison versus various tools.
    Key: S/s=safe, U/u=unsafe, B/b=bug-detected, N/n=no-bug-detected,
         C/c=crash, t.o.=time-out, b.r.=bound-reached,
         uppercase/bold=positive-result,
         lowercase=negative-result.
    \label{fig:tools}}
\end{figure*}

For our experiments, we compare variants of
a simple \emph{safe program} 
consisting of the following basic template:
\begin{verbatim}
    list_node *xs = make_list(n); set(xs, m, v);
\end{verbatim}
Here the \verb+set+ function is defined in Section~\ref{sec:closed}, and
\verb+make_list+ constructs a linked-list of length $n$.
We also evaluate several unsafe variants of the basic template, including:
\begin{itemize}
    \item[-] \verb+overlap_node+: list with overlapping nodes, e.g.,
        Figure~\ref{fig:lists}(g);
    \item[-] \verb+wrong_node+: using the wrong node type, e.g.
        \verb+list_node+ for a \verb+tree_node+ function;
    \item[-] \verb+wrong_size+: passing the wrong size to \verb+malloc+;
    \item[-] \verb+not_array+: attempt to access a list element via an
        array subscript, e.g. \verb+xs[1].next+;
    \item[-] \verb+cast_int+: manufacture an invalid pointer from an
        integer, e.g. $(\mathtt{list\_node}~*)i$ for integer $i$;
    \item[-] \verb+uninit_ptr+: neglecting to initialize a pointer, e.g.
        Figure~\ref{fig:lists}(f);
    \item[-] \verb+uninit_ptr_stk+: neglecting to initialize a pointer on the
        stack; and
    \item[-] \verb+arith_ptr+: arbitrary pointer arithmetic, e.g.
        \verb+(xs-3)->next+.
\end{itemize}
Each unsafe variant is \emph{exploitable} in that it demonstrably
(by compiling and running the program) overwrites
memory outside of the footprint.
In addition to bounded linked-lists (\emph{List}), we also test a
variant that uses parameterized \emph{DAG} and a
parameterized graph \emph{Graph} in place of lists.

The experimental results are shown in Figure~\ref{fig:tools} with the
result key summarized by the caption.
In the ideal case, we expect the following: static analysis tools should 
report $\{\text{S}, \text{U}\}$;
BMC-tools should report $\{\text{B}, \text{N}\}$; and
dynamic bounds checking tools should report
$\{\text{B}, \text{C}, \text{N}\}$.
All tools were fast (${<}10s$) provided no timeout/bounds-reached
condition occurred.
Our experimental comparison excludes
\textsc{SmallFoot} (no C support),
\textsc{SLAyer} (crashed with error),
Verifast (requires manual proofs), and
LowFat for \emph{DAG} and \emph{Graph}
(bug not reachable).
The $\DD$-tool 
performs as expected (total score 27/27) for all data-structure shapes.
The results for \textsc{SpaceInvader} were mixed, even for lists.
In contrast, Predator performed flawlessly for 
list data-structures but less well in
the \emph{DAG} and \emph{Graph} tests.
For the latter, Predator
appears to resort to (infinite) unfolding leading to timeouts.
The results for the BMC-based tools are also generally positive.
CBMC and LLBMC detect most memory errors for the unsafe test
cases, demonstrating that the BMC approach is effective
at detecting bugs.
The total scores are LLBMC (22/27) and CMBC (21/27).
There are also some anomalous results, e.g.,
SpaceInvader reports unsafe programs as safe.
CBMC reports a false null-pointer error for the
\emph{DAG}/\verb+overlap_node+ test case.
LowFat, as a bounds checker, primarily detects errors relating
to bad pointer arithmetic, and may not detect memory errors
relating to bad casts (type confusion) or uninitialized pointers.
We highlight that, while different tools embody different tradeoffs
(Figure~\ref{fig:toolcomp}), the $\DD$-tool focuses on general 
data structures, modularity and (limited) memory safety.

%%%%%%%%%%%%%%%%%%%%%%%%%%%%%%%%%%%%%%%%%%%%%%%%%%%%%%%%%%%%%%%%%%%%%%%%%%%%%
\section{Conclusion}\label{sec:conclude}

This paper presented a shape neutral data-structure analysis for 
low-level heap manipulating programs.
The analysis validates several key properties of graph-based data-structures
including the validity of nodes, pointers, and the separation between nodes.
Such properties are standard for graph-based data-structures implemented
in idiomatic C.
Our approach therefore caters for a broad range of heap-manipulating
code.

Our analysis methodology is based on using symbolic execution to generate
Verification Conditions (VCs) which
are then solved using a specialized data-structure property solver
(a.k.a., the $\DD$-solver) in combination with built-in
heap, integer and equality solvers.
The $\DD$-solver itself is implemented using Constraint Handling Rules (CHR),
and is automatically generated from the type declarations contained
within the target program.
For solving VCs, our $\DD$-tool employs the
\emph{Satisfiability Modulo Constraint Handling Rules} (SMCHR) system.
The SMCHR system is well suited for this task, as it can handle VCs
with a rich Boolean structure, rewrite rules (for negation), CHR solvers
(for the $\DD$-solver), and allows
for the integration of different kinds of solvers
(i.e., integer, heap and the $\DD$-solver).
Our experimental results are promising, with the $\DD$-tool able to
detect memory errors that are missed by other tools.

\section*{Acknowledgements}
This research was partially supported by MOE2015-T2-1-117 and
R-252-000-598-592.

\bibliographystyle{acmtrans}
\bibliography{unbox}

\end{document}